\documentclass[superscriptaddress,preprintnumbers,showpacs,nofootinbib]{revtex4}
\usepackage{amssymb,graphicx,dcolumn,bm}

\begin{document}

\newcommand*{\PKU}{School of Physics and State Key Laboratory of Nuclear Physics and
Technology, \\Peking University, Beijing 100871}\affiliation{\PKU}
\newcommand*{\CHEP}{Center for High Energy
Physics, Peking University, Beijing 100871}\affiliation{\CHEP}

\title{Lorentz violation effects on astrophysical propagation of very high energy photons}

\author{Lijing Shao}\affiliation{\PKU}
\author{Bo-Qiang Ma}\email{mabq@pku.edu.cn}\affiliation{\PKU}\affiliation{\CHEP}

\begin{abstract}
Lorentz violation (LV) is predicted by some quantum gravity (QG)
candidates, wherein the canonical energy-momentum dispersion
relation, $E^2=p^2+m^2$, is modified. Consequently, new phenomena
beyond the standard model are predicted. Especially, the presence of
LV highly affects the propagation of astrophysical photons with very
high energies from distant galaxies. In this paper, we review the
updating theoretical and experimental results on this topic. We
classify the effects into three categories: (i) time lags between
photons with different energies; (ii) a cutoff of photon flux above
the threshold energy of photon decay, $\gamma \rightarrow e^++e^-$;
(iii) new patterns in the spectra of multi-TeV photons and EeV
photons, due to the absorption of background lights. As we can see,
the details of LV effects on astrophysical photons depend heavily on
the ``phase space'' of LV parameters. From observational aspects,
available and upcoming instruments can study these phenomena
hopefully, and shed light onto LV issues and QG theories. The most
recent progresses and constraints on the ultra-high energy cosmic
rays (UHECRs) are also discussed.
\end{abstract}


\maketitle

\section{Introduction}

\label{sec1}

The unification of standard model and general relativity is one of
the most intriguing and desirous goals of modern physics, and it has
stimulated many theoretical ideas towards quantum gravity (QG). Some
of them predict Lorentz violation (LV), where the Lorentz symmetry
of space-time breaks down at high energies, and it introduces some
tiny ``LV relics'' at lower energies. This probability has arisen in
the space-time foam~\cite{a97,a98,emn00,emn00b,emn08,emn09}, loop
gravity~\cite{gp99,amu00,alfaro03}, vacuum condensate of 
tensor fields~\cite{ck97,ck98}, and the so-called doubly special
relativity (DSR; also known as the deformed special
relativity)~\cite{a02,a10}.

LV is an important issue in QG theories, mainly because it provides
some available scenarios to validate or falsify theories through
``windows on QG''~\cite{jlm06,lm09}. Generally, we know that the
quantum-gravitational effects are relevant only when the energy
approaches the Planck energy, $E_{\rm Pl} \equiv \sqrt{\hbar c^5 /G}
\simeq 1.22 \times 10^{19}$~GeV.\footnote{There are also arguments
on a new energy scale rather than the Planck energy as the quantum
gravity scale, see {\it e.g.}, Ref.~\cite{sm10} and references
therein.} This huge number is obviously out of reach in the
foreseeable future. However, at lower energies, the ``relic''
effects can demonstrate themselves and produce a modified
energy-momentum dispersion relation with extra terms suppressed by
$\xi_n(E/E_{\rm Pl})^n$. Here the power $n$ and the coefficient
$\xi_n$ both depend on the specific theoretical framework under
consideration~\cite{jlm06,lm09,m05}.

To see the manifestations of the Lorentz-violating physics, novel
insights are needed, and fortunately, some have already been
proposed practically to probe tiny modifications to conventional
physics. Accurate measurements in the laboratories on the
earth~\cite{a09,muller07}, violating processes in astrophysics with
energetic
particles~\cite{emn09,pm00,s03,jlm03,gm04,gs08,gs08b,jp08,m09,b09,mls10,alfaro05,alfaro05b,xm09b,mu09,bi08},
and accumulated effects through cosmological amplification
mechanisms~\cite{a98,emn00,emn08,emn09,e99,e06,jp08b,a08,xm09,fermi09a,fermi09b,as09,ss09,sxm10},
are three basic considerations till now to reveal new physics arisen
in the LV scenarios.

Here we try to review and discuss some candidate theories on LV, and
then utilize the modified energy-momentum dispersion relation to
search observational clues in the photon sector (sometimes jointly,
the electron and positron sectors). High energy photons originated
from astrophysics are extremely promising to probe or falsify LV
theories in various aspects~\cite{jlm06,lm09,m05}. This paper is
organized as follows. In Sec.~\ref{sec2}, several selected
underlying models are briefly reviewed and discussed with the most
recent observational constraints. In Sec.~\ref{sec3}, we focus on
photons from astrophysical sources. These photons can reveal LV
effects through their time-of-flight with energy-dependent
velocities, and their interactions with low energy background
photons, namely the extra-galactic background light (EBL) and/or the
$2.7$ K cosmic microwave background (CMB) radiations. Multi-TeV
$\gamma$-rays may also be involved in self-decays into $e^+e^-$
pairs with proper LV parameters. In Sec.~\ref{sec4}, the most
updating LV constraints from the spectra of ultra-high energy cosmic
rays (UHECRs) are reviewed and discussed. Sec.~\ref{sec5} summaries
the paper with discussions on several current and upcoming
instruments related to LV physics.

\section{Some Selected Theoretical Models}

\label{sec2}

In a rough manner, LV theories can mainly be classified into two
categories. One is effective field theory (EFT), which provides an
excellent framework where tiny LV effects are introduced into the
Lagrangian through LV operators --- renormalizable ones with mass
dimension 3 and/or 4, see, {\it e.g.}, the standard model extension
(SME)~\cite{ck97,ck98}, and the further extended non-renormalizable
ones with mass dimension 5 and/or
6~\cite{mp03,m08,km07,km08,km09,au10}. The other category includes
those QG theories that cannot be embedded into the EFT framework,
such as the quantum space-time foam
model~\cite{a97,a98,emn00,emn00b,emn08,emn09} and
DSR~\cite{a02,a10}.

In the EFT framework, all renormalizable operators, which guarantee
the gauge symmetry preserved, are introduced by Colladay and
Kosteleck\'y~\cite{ck97,ck98} perturbatively and elegantly. The LV
terms can be characterized by dimensionless tiny coefficients of
order $\mathcal{O} (10^{-23})$ or even smaller. Lately, Myers and
Pospelov~\cite{mp03} proposed dimension-5 non-renormalizable LV
terms into the EFT Lagrangian. The dimension-5 CPT-odd terms in
quantum electrodynamics (QED) read as~\cite{mp03,m08},
\begin{equation}
{\cal L}_{\rm LV^-}^{\rm 5d,\,QED} = \frac{1}{E_{\rm Pl}} \xi^{(5)}
u^\mu u_\rho F_{\mu\nu} (u \cdot
\partial) \tilde{F}^{\rho\nu}
+ \frac{1}{E_{\rm Pl}} \bar{\psi} (\delta_L^{(5)} P_L +
\delta_R^{(5)}P_R ) (u \cdot \gamma) (u \cdot D)^2 \psi \,,
\end{equation}
while the dimension-5 CPT-even terms appear as~\cite{mp03,m08}
\begin{equation}
{\cal L}_{\rm LV^+}^{\rm 5d,\,QED} = - \frac{1}{E_{\rm Pl}}
\bar{\psi} (u \cdot D)^2 ( \zeta^{(5)}_L P_L + \zeta^{(5)}_R P_R)
\psi \,.
\end{equation}
In above Lagrangians, $u$ is a fixed timelike four-vector indicating
the preferred frame, which is usually chosen as the frame where the
CMB radiation is isotropic; $\tilde{F}$ is the dual of $F$, defined
as $\tilde{F}^{\mu\nu} = \epsilon^{\mu\nu\rho\lambda}
F_{\rho\lambda} /2$; $P_{R,L} = (1\pm \gamma^5)/2$ are projection
operators, and $D_\mu = \partial_\mu + i e A_\mu$ is the QED
covariant derivative; $\xi$, $\delta_{R,L}$, and $\zeta_{R,L}$ are
dimensionless LV coefficients naturally expected to be of order
$\mathcal{O} (1)$. Lately, the dimension-6 CPT-even terms are also
introduced, by Mattingly~\cite{m08},
\begin{eqnarray}
{\cal L}_{\rm LV^+}^{\rm 6d,\,QED} & = & - \frac{1}{2E_{\rm Pl}^2}
\xi^{(6)} u^\mu u_\rho F_{\mu\nu} (u \cdot
\partial)^2 {F}^{\rho\nu} \\ \nonumber
& &- \frac{i}{E_{\rm Pl}^2} \bar{\psi} (u \cdot D)^3 (u \cdot
\gamma) ( \zeta^{(6)}_L P_L + \zeta^{(6)}_R P_R) \psi \\ \nonumber%
& & - \frac{i}{E_{\rm Pl}^2} \bar{\psi} (u \cdot D) \square (u \cdot
\gamma) ( \bar{\zeta}^{(6)}_L P_L + \bar{\zeta}^{(6)}_R P_R) \psi
\,.
\end{eqnarray}

We can see that in the non-renormalizable EFT, the LV terms are
spontaneously suppressed by $(E/E_{\rm Pl})$ or $(E/E_{\rm Pl})^2$,
compared to the renormalizable LV terms with tiny coefficients of
order $\mathcal{O} (10^{-23})$ put in by hand~\cite{ck97,ck98,cg99}.
However, there remains a ``naturalness
problem''~\cite{lm09,c04,alfaro10} in the non-renormalizable EFT ---
radiative corrections may generate ``baby'' renormalizable terms
which subsequently become even more important than the original
``parent'' non-renormalizable operators. Luckily, supersymmetry can
play a role to prevent or alleviate this problem~\cite{np05,bnp05}.

From the above Lagrangians, we finally arrive at the modified
energy-momentum dispersion relations for photons and fermions,
respectively,
\begin{equation}
\omega^2 = k^2 + \xi^{(n)}_{\pm} \frac{k^{n-2}}{E_{\rm Pl}^{n-4}}
\,,
\end{equation}
\begin{equation}
E^2 = p^2 + m^2 + \eta_{\pm}^{(n)} \frac{p^{n-2}}{E_{\rm Pl}^{n-4}}
\,,
\end{equation}
where $m$ is the rest mass of fermions, and $\eta_{\pm}$ are
combinations of $\delta_{R,L}$ and/or $\zeta_{R,L}
(\bar{\zeta}_{R,L})$~\cite{lm09,mp03,km07,km08,km09}, where their
subscribes ``$\pm$'' denote two opposite helicities. In the EFT,
$\xi_+^{(n)} = (-)^n \cdot \xi_-^{(n)}$, and $\eta^{(n)}_{\pm}
\{e^+\} = (-)^n \cdot \eta^{(n)}_{\mp} \{e^-\}$~\cite{lm09}.
However, $\eta_+$ is not necessary to equal to $\eta_-$ in general.
For more details on non-renormalizable EFT calculations and related
issues, see {\it e.g.}, Refs.~\cite{lm09,m08,km07,km08,km09}.

From aspects of observational limits, unfortunately, the dimension-5
operators seem unnatural, even with supersymmetry included and
softly broken~\cite{np05,bnp05}. The linear energy dependence in EFT
is largely unfavored~\cite{jlm06,lm09,m05,kr08}. The most strict
observations come from the vacuum birefringence and synchrotron
radiations~\cite{jlm06,lm09,m05}. So we would like to fix $n=6$ in
the following analysis, and impose $\eta_+ = \eta_-$ for simplicity
(denoted as $\eta$ hereafter; Liberati and Maccione also treated EFT
in the same way in Ref.~\cite{lm09}). This choice preserves CPT
symmetry and helicity symmetry, though they are not the essential
ingredients generally.

There are also several models where the local description in terms
of effective Lagrangian breaks down. For instance, Ellis {\it et
al.} proposed a Liouville-inspired stringy analogue of space-time
foam model~\cite{a97,a98,emn00,emn00b,emn08,emn09}. In this model, a
gas of D-particles roams in the bulk space-time of a high-dimension
cosmology, wherein our universe is represented as a D3-brane.
Photons are represented by open strings. Their interaction with
D-particles excites and stretches the strings, and then decays and
emits photons. Thus ``gravitational medium effects'' modify the
canonical dispersion relation of photons to include an LV term which
depends linearly on the energies of photons, $\omega^2 = k^2[1 + \xi
(k / E_{\rm Pl})]$, while the dispersion relation for charged
particles remains untouched and no helicity dependence is
predicted~\cite{a97,a98,emn00,emn00b,emn08,emn09}. Hence, it avoids
many tight constraints from the vacuum birefringence of photons, the
electron/positron sector, and the UHECRs
sector~\cite{mu09,bi08,kr08}. Most recently, Ref.~\cite{mls10}
claimed that the space-time foam model is unable to explain
simultaneously the time lags of TeV $\gamma$-rays, and the
non-observation of high energy photons above EeV. However, Ellis,
Mavromatos, and Nanopoulos argued that there are ways to avoid the
constraints~\cite{emn10}, hence the issue is still in dispute. On
the other hand, the time-of-flight of photons from distant
$\gamma$-ray bursts (GRBs) and active galactic nuclei (AGNs)
strikingly favors the linearly energy dependent LV
corrections~\cite{emn08,emn09,sxm10}. Therefore, it still has strong
motivations and advantages to study this linearly dependent LV
theory.

The above discussed LV theories are far from complete. Especially,
DSR~\cite{a02,a10} is a well-motivated candidate where two invariant
scales are included, namely the velocity of low energy photons, {\it
i.e.}, the conventional light speed $c$, and a length scale $l_{\rm
Pl} \equiv \sqrt{G\hbar /c^3} \simeq 1.61 \times 10^{-35}~{\rm m}$
(or equivalently, an energy scale $E_{\rm Pl}$). In the DSR, the
energy-momentum conservation laws may also be modified~\cite{a10},
thus brings further complications to the analysis of LV phenomena.
In this paper, the canonical energy-momentum conservation is
assumed.

For a phenomenologically uniform description, we use the following
dispersion relations, for photons and fermions, respectively,
\begin{equation}\label{photon}
\omega^2 = k^2 \left[1 + \xi_{n} \left(\frac{k}{E_{\rm Pl}}\right)^n
\right] \,,
\end{equation}
\begin{equation}\label{fermion}
E^2 = m^2 + p^2 \left[ 1+ \eta_{n} \left(\frac{p}{E_{\rm
Pl}}\right)^n \right]\,,
\end{equation}
where only $n=1$ (the linear modification) and $n=2$ (the quadratic
modification) are currently observationally relevant. More
specifically, we should keep in mind that the case with $\xi_1 \neq
0$, $ \eta_1 = \eta_2 = \xi_2 =0$ corresponds to the space-time foam
model, while the case with $\xi_1=\eta_1=0$, $\xi_2 \neq 0$, $\eta_2
\neq 0$ represents the dimension-6 EFT with CPT-even LV operators.
In addition, the case with $\xi_1=\eta_1 \neq 0$, $\xi_2=\eta_2=0$
is also of great interests, for the reason that it can be a
consequence of the geometrically small scale of space-time and the
equivalence principle to all species of particles.

\section{Astrophysical Photons in Presence of LV}

\label{sec3}

As the dispersion relations are modified, we can suspect extra LV
impacts beyond the standard model. The tiny corrections can be
magnified when the energy $E$ is large, and also maybe through
cosmological amplification mechanisms. Hence when two magnification
mechanisms are combined, the astrophysical photons with very high
energies provide plentiful motivations to be regarded as an
important source to investigate LV physics.

The dynamics of LV is poorly understood, hence we focus on the
kinematics of photon propagation from distant galaxies towards us.
The effect of LV on high energy photon propagation is threefold.
Firstly, the velocity of photons, defined as $v = \partial \omega /
\partial k$, is no longer a constant, according to
Eq.~(\ref{photon}). Consequently, photons with different energies
spend slightly different time-of-flight to arrive at the earth. It
was firstly recognized in Refs.~\cite{a97,a98}, and is discussed
extensively by the scientific society with great passion recently on
the Fermi, Magic, and HESS
observations~\cite{xm09,fermi09a,fermi09b,as09,sxm10}. Secondly, the
modified dispersion relation may allow the strictly forbidden
reactions in the Lorentz-symmetric theories to occur, such as photon
decay, $\gamma \rightarrow e^- + e^+$. It results in a cutoff in the
spectra of $\gamma$-rays above the decay threshold. Thirdly, high
energy photons interact with low energy background photons (CMB
and/or EBL) on the way of propagation, through the reaction $\gamma
+ \gamma_{\rm CMB/EBL} \rightarrow e^+ + e^-$; LV effects can
influence the threshold of this process, and lead to interesting
phenomena~\cite{pm00,s03,jlm03,gs08,gs08b,jp08,mls10}.

We discuss the above three scenarios in the following. Worthy to
stress that, since the Lorentz symmetry is broken in the LV theory,
the transformation laws between two relatively moving frames are
unknown, hence the laboratory system is the only proper frame to
discuss relevant physics before we get access to proper
transformation laws.

\subsection{Lags from time-of-flight}

\label{sec3a}
%

Amelino-Camelia {\it et al.}~\cite{a97,a98} first suggested using
GRBs to test LV physics.  Due to the large cosmological distance and
the fine time structure of GRBs, tiny LV effects can be amplified
into observable quantities. By taking into account cosmological
expansion of the universe, the time lag induced by the LV modified
dispersion relation between photons with high energies $E_{\rm h}$,
and those with low energies $E_{\rm l}$, is~\cite{jp08b},
\begin{equation}\label{lag}
\Delta t_{\rm LV} = \frac{1+n}{2 H_0} \xi_n \left( \frac{E_{\rm l}^n
- E_{\rm h}^n}{E_{\rm Pl}^n} \right) \int_0^z \frac{(1+z^\prime)^n
{\rm d} z^\prime}{\sqrt{\Omega_{\rm \Lambda}  + \Omega_{\rm M}
(1+z^\prime)^3 }} \,,
\end{equation}
where $n=1$ and $n=2$ stand for linear and quadratic energy
dependence; $H_0 \simeq 71$~km~s$^{-1}$~Mpc$^{-1}$ is the Hubble
expansion constant; $\Omega_{\rm \Lambda} \simeq 0.73$ is the vacuum
energy density, and $\Omega_{\rm M} \simeq 0.27$ is the matter
energy density in the current universe; $z$ is the redshift of the
source.

Fermi LAT discovered that high energy photons have a tendency to
arrive later relative to low energy ones~\cite{fermi09a,fermi09b},
which might present potential evidence for LV with $\xi_1 <0$ for
the space-time foam model or $\xi_2 <0$ for the dimension-6 CPT-even
EFT. However, the determination of time lag from observational data
is highly nontrivial and affected by many facets, both artificially
and instrumentally~\cite{fermi09b,sxm10}. The primary uncertainty
comes from the unknown effects from source activities, mainly due to
our imperfect knowledge of the radiation mechanism of GRBs. However,
we can separate the source effects if we can achieve a survey of
GRBs at different redshifts~\cite{e99,e06,sxm10}. The time lag
induced by LV accumulates with propagation distance, as it is a
gravitational medium effect. On the contrary, the intrinsic source
induced lag is likely to be a distance-independent quantity, which
can be approximated as a constant. The relation between the observed
delay $\Delta t_{\rm obs}$ and the intrinsic time lag $\Delta t_{\rm
in}$ is
\begin{equation}\label{tlag}
\Delta t_{\rm obs} = \Delta t_{\rm LV} + \Delta t_{\rm in} (1+z) \,.
\end{equation}
In Ref.~\cite{sxm10}, we systematically studied this scenario of
Fermi LAT GRBs, and got the quantum gravity scale $E_{\rm QG} \sim 2
\times 10^{17}$~GeV for the linear energy dependence, which
corresponds to $\xi_1 \sim -10^2$. Worthy to mention that, $\xi_n
> 0$ represents superluminal photon propagation, while $\xi_n < 0$
corresponds to subluminal propagation.

The time-of-flight of high energy photons is the clearest test to
search LV effects of the photon sector, hence provides the most
convincing results. AGNs and pulsars are also utilized to study LV
effects through time-of-flight analysis~\cite{jlm06,lm09,m05,sxm10}.
The observational results that high energy photons arrive relatively
later than low energy ones suggest a negative $\xi_n$, and current
data largely favor linearly energy dependent LV
effects~\cite{as09,sxm10}.

\subsection{Photon decay: a cutoff in the $\gamma$-ray spectra}

\label{sec3b}

The alteration of the dispersion relations of photons and fermions
might also lead to some peculiar reactions to occur, like the basic
QED vertex $\gamma \rightarrow e^++e^-$, called photon decay, which
is generally forbidden by the canonical energy-momentum conservation
in the standard model.

Photon decay might be realistic in the framework of LV theories with
proper LV parameters. We consider a very high energy photon of
momentum $k$, which decays into an electron of momentum $xk$, $x \in
[0,1]$, and a positron of momentum $(1-x)k$. Utilizing the
``threshold theorem''~\cite{jlm03b}, energy conservation leads
to~\cite{jlm03}
\begin{equation}
k\left[1 + \frac{\xi_{n}}{2} \left(\frac{k}{E_{\rm Pl}}\right)^n
\right] = xk \left[ 1+ \frac{m^2}{2(xk)^2}+\frac{\eta_{n}}{2}
\left(\frac{xk}{E_{\rm Pl}}\right)^n \right] + \{x \leftrightarrow
1-x\} \,,
\end{equation}
where we only keep the leading corrections of Planck-scale terms,
and the leading term in the series of $(m/k)^2$. The consistence of
the truncation can be checked. After a few steps, the equation turns
into~\cite{jlm03}
\begin{equation}\label{pd}
\frac{m^2E_{\rm Pl}^n}{k^{n+2}} = x(1-x) 
\left[ \xi_n - \eta_n \left((1-x)^{n+1}+x^{n+1}\right) \right] .
\end{equation}
To get the threshold to valid photon decay, is equivalent to
minimize $k$ on the left-hand side, hence to maximize the right-hand
side of Eq.~(\ref{pd})~\cite{jlm03}. We consider three theoretical
scenarios in the following.

(i) $\mathbf{\xi_1 \neq 0,  \eta_1 = \eta_2 = \xi_2 =0}$

This is the LV parameter configuration for the space-time foam
model~\cite{a97,a98,emn00,emn00b,emn08,emn09}. Under these
circumstances, Eq.~(\ref{pd}) reduces to $m^2E_{\rm Pl} /k^{3} =
[x(1-x)] \xi_1$. We can see clearly that $\xi_1 \rightarrow 0$ leads
to $k \rightarrow +\infty$, which is the very case in the standard
model where no photon decay is kinematically allowed. For $\xi_1 <
0$, there is no meaningful solution for the photon decay process. On
the contrary, for $\xi_1>0$, any high energy photon with its
momentum larger than $k_{\rm th} = (4m^2E_{\rm Pl}/\xi_1)^{1/3}
\simeq 23.4 \, {\xi_1}^{-1/3}~{\rm TeV} $ would decay rapidly into
an $e^+e^-$ pair~\cite{jlm03}, thus we are not supposed to observe
photons with momentum larger than $23.4 \, {\xi_1}^{-1/3}~{\rm TeV}$
from astrophysical sources.

On the other hand, the observation of multi-TeV photons can cast a
constraint on the LV parameter $\xi_1$ from the $\xi_1>0$
side~\cite{s03}. If $\xi_1$ is of order $\mathcal{O}(1)$, then the
relevant energy lays within the currently observational capability.
Actually, the observations of the $50~{\rm TeV}$ and $80~{\rm TeV}$
photons from the Crab Nebula~\cite{lm09} have already constrained
$\xi_1$ to be less than $10^{-2}$ in this scenario. It should be
noticed that, here the power dependence of $k_{\rm th}$ on $\xi_1$,
$-1/3$, differs from that of Ref.~\cite{s03}, $-1/2$, because there
the energy dependence of LV in terms of the maximal attainable
velocity~\cite{cg99} is different from that of the space-time foam
model. However, in the relevant energy range, the results are
roughly the same.

(ii) $\mathbf{\xi_1=\eta_1 \neq 0, \xi_2=\eta_2=0}$

If the modification of the dispersion relation is due to pure
Planck-scale geometry of space-time, and the equivalence principle
applies to all species of particles, then the fermions (electrons
and positrons) share the same LV parameters as the photons. And if
the dispersion relation is modified to the first order, {\it i.e.},
linearly energy dependent, then from Eq.~(\ref{pd}), the
energy-momentum conservation leads to $m^2E_{\rm Pl} /k^{3} =
2[x^2(1-x)^2] \xi_1$. Thus again, for $\xi_1=\eta_1 \leq 0$, no
photon decay is allowed, and for $\xi_1=\eta_1>0$, there exists a
threshold for photons with momentum larger than $k_{\rm th} =
(8m^2E_{\rm Pl}/\xi_1)^{1/3} \simeq 29.4 \, {\xi_1}^{-1/3}~{\rm TeV}
$ to decay. This is also within the energy range of observational
practicability if the LV parameters are of order $\mathcal{O}(1)$.

(iii) $\mathbf{\xi_1=\eta_1=0, \xi_2 \neq 0, \eta_2 \neq 0}$

In the above two cases, the threshold happens when $x=1/2$, {\it
i.e.}, the electron and the positron in the final state have the
same amount of momentum. However, this is not a general property in
the LV photon decay phenomenon, as first pointed out by Jacobson,
Liberati and Mattingly~\cite{jlm03}.

In the dimension-6 CPT-even EFT framework, LV terms of photons and
fermions both depend quadratically on the energy. Now Eq.~(\ref{pd})
is reduced to $m^2 E_{\rm Pl}^2 /k^{4} = [x(1-x)] \times \left[
\xi_2 - \eta_2 \left((1-x)^{3}+x^{3}\right) \right]$, where the
right-hand side is quartic of $x$ if $\eta_2 \neq 0$, and the
discussions to derive the threshold energy become more subtle. If
$\eta_2 = 0$, then it is a similar case as the space-time foam
model, where $\xi_2 \leq 0 $ corresponds to no photon decay, while
$\xi_2
>0$ leads to a threshold at $k_{\rm th} = (4m^2E_{\rm
Pl}^2/\xi_2)^{1/4} \simeq 0.11~\xi_2^{-1/4}~{\rm EeV}$.

\begin{figure}
\begin{center}
\includegraphics[width=7.0cm]{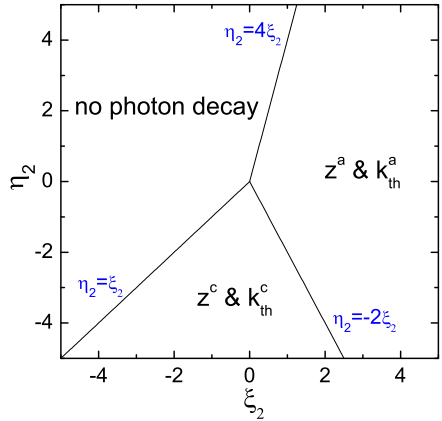}
\caption{The ``phase diagram'' of photon decay in the dimension-6
CPT-even EFT with LV parameters $\xi_2$ and $\eta_2$ in arbitrary
unit. \label{pdiagram}}
\end{center}
\end{figure}

For the $\eta_2 \neq 0$ case, after introducing $z = (2x-1)^2$,
$z\in[0,1]$, the constraint becomes quadratic of $z$~\cite{jlm03},
\begin{equation}
\frac{m^2E_{\rm Pl}^2}{k^4} = %
\frac{3\eta_2}{16} \left(z - \frac{2\xi_2 + \eta_2}{3\eta_2}
\right)^2 - \frac{1}{12\eta_2} (\xi_2-\eta_2)^2 \,.
\end{equation}
The maximum of the right-hand side can happen at $z^a=0$ (no
asymmetry), $z^b=1$ (the maximum asymmetry), or $z^c=(2\xi_2 +
\eta_2)/3\eta_2$ (a mediate asymmetry), depending on the realistic
values of LV parameters~\cite{jlm03}. After a detailed calculation,
we can get the ``phase diagram'' for photon decay in the dimension-6
CPT-even EFT, illustrated in Fig.~\ref{pdiagram}~\cite{jlm03}, where
\begin{equation}
k_{\rm th}^{\rm a} = \left( \frac{16m^2E_{\rm
Pl}^2}{4\xi_2-\eta_2}\right)^{1/4} \simeq 0.16
\,(4\xi_2-\eta_2)^{-1/4}~{\rm EeV} \,,
\end{equation}
\begin{equation}
k_{\rm th}^{\rm c} = \left[ \frac{-12 \eta_2 m^2E_{\rm
Pl}^2}{(\xi_2-\eta_2)^2} \right]^{1/4} \simeq 0.15 \left[
\frac{-\eta_2}{(\xi_2-\eta_2)^2} \right]^{1/4}~{\rm EeV} \,.
\end{equation}

We can see that the physical process $\gamma\rightarrow e^++e^-$
heavily depends on the phase space of LV parameters. In the region
where $z=z^a=0$ ($x=1/2$ thereof), it appears the naturally expected
configuration of final state, where the electron and the positron
share the same amount of the initial momentum. However, in some
other phase space, the asymmetric momentum configuration of final
state may appear, namely that the electron and the positron gain
different proportions of the initial momentum, satisfying
$z=z^c=(2\xi_2 + \eta_2)/3\eta_2$ ($x=[1\pm
\sqrt{(2\xi_2+\eta_2)/3\eta_2}~]/2 \neq 1/2$ thereof)~\cite{jlm03}.
Besides, there is also a regime where no photon decays can ever
happen.

From the above discussions, we can see that within different
theories, there can be a cutoff in the $\gamma$-ray spectra above a
threshold energy with LV parameters in some certain regions. The
cutoff is predominant, instead of a suppression, because the time
scale of the photon decay is quite short and the process takes place
in no time compared to their
time-of-flight~\cite{jlm06,s03,cg99,ks08}. Thus, the photon decay
feature is obvious in the photon spectra if it exists, and hence
should be observed explicitly. The $n=2$ case has the cutoff energy
four orders higher than that of the $n=1$ case, thus it is more
challenging to practical observations. However, the observations
concerning the quadratic LV modification, as well as the linear LV
modification, are already emerging.

\subsection{Modifications from pair-production absorption: an enhancement in the $\gamma$-ray spectra}

\label{sec3c}

Another interesting phenomenon comes from the interaction of a high
energy photon $\gamma$ with a low energy photon $\gamma_{\rm
CMB/EBL}$ through the pair-production reaction $\gamma + \gamma_{\rm
CMB/EBL} \rightarrow e^++e^-$. CMB is the relic radiation from the
big bang which evolutes in the expanding universe after the lights
decouple with the matters. And EBL is the faint diffuse light of the
sky, consisting of the historically accumulated flux of all
extragalactic sources. It is relevant for the formation of stars and
galaxies, and also the large-scale structure of the universe. Its
wavelength varies from the radio band to the ultraviolet band. The
observed spectra of high energy photons from distant galaxies are
modified on the way of propagation, due to the attenuation by these
low energy background ``absorbers''.

The presence of LV will change the reaction threshold and modify the
reaction patterns in an intriguing and physically meaningful
manner~\cite{s03,jp08}. It can be extended to the ultra-high energy
photons, as well as the UHECRs, without any further
difficulties~\cite{gs08,gs08b,m09,b09,mls10,alfaro05,alfaro05b,xm09b,bi08,ss09}.
This might be the most extensively investigated scenario among the
LV-induced ``windows on QG''~\cite{jlm06,lm09,m05}. The LV effects
depend on the energy of photons involved, and the magnitude and also
the sign of the LV parameters, as well as the theoretical framework
under considerations.

As we can see in the photon decay section, different parameter
configurations lead to different LV physical phenomena. Since four
particles are involved in the pair-production absorption, we
naturally expect more subtle and complicated dependence on the phase
space of LV parameters than the photon decay case. Jacobson {\it et
al.} provided the most detailed discussions on this
issue~\cite{jlm06,jlm03}. Here we look into two possible scenarios
which are the most relevant and well constrained from observational
aspects: (i) the propagation of multi-TeV photons~\cite{s03,jp08},
(ii) the photo-meson production through the GZK mechanism and the
subsequently pion-produced photons with energies above
EeV~\cite{gs08,gs08b,mls10}.

We would work in the simple framework provided by Jacob and
Piran~\cite{jp08}, who studied the linearly LV suppressed scenarios
in a clear manner. Assuming that the LV parameters of fermions
vanish, then by utilizing the ``threshold theorem''~\cite{jlm03b},
we can arrive at the threshold of the low energy background lights
to interact with an incidental photon with energy
$\omega$~\cite{jp08},
\begin{equation}\label{abeq}
\epsilon_{\rm th} = \frac{m^2}{\omega} - \frac{\xi_n}{4}
\left(\frac{\omega}{E_{\rm Pl}}\right)^n \omega \,.
\end{equation}
$\xi_n =0$ leads to the conventional case $\epsilon_{\rm th} =
m^2/\omega$, of course. If the LV parameters of electrons and
positrons are also introduced, the LV modified patterns are hardly
further modified. It only introduces a numerical rescaling of order
$\mathcal {O} (1)$ for the second term in Eq.~(\ref{abeq}), by
assuming the likely magnitude of the LV parameters of fermions
compared to that of photons~\cite{jp08}. Thus we here ignore them
for simplicity.

In the superluminal propagation case with $\xi_n >0$, the photon
decay process dominates, where it results in a cutoff, and no
photons are observable above energy larger than the decay threshold.
Thus we focus on the subluminal case with $\xi_n <0$ here. It is
easily seen that there is a minimum on the right hand side of
Eq.~(\ref{abeq}). The minimum appears as the most crucial feature
from the pair-production absorption within LV theories~\cite{jp08}.
Firstly, it means that the background photons with energies lower
than the minimum never interact with the high energy photons.
Secondly, most importantly, LV predicts a ``reemergence'' of the
high energy photon flux when the incidental photons have energies
above the critical energy~\cite{jp08},
\begin{equation}
\omega_{\rm cr} = \left[ -\frac{4  m^2 E_{\rm Pl}^n}{(n+1)\xi_n}
\right]^{1/(n+2)} \,.
\end{equation}
For the $n=1$ case, $\omega_{\rm cr} \simeq 18.5
\,(-\xi_1)^{-1/3}~{\rm TeV} $, while for the $n=2$ case,
$\omega_{\rm cr} \simeq 84.8 \,(-\xi_2)^{-1/4}~{\rm PeV} $. The
critical energies are of the same order as the threshold energies of
photon decays, because the photon decay can be viewed as a high
energy $\gamma$-ray photon interacting with a ``low'' energy one
with infinite wavelength (therefore zero energy actually). Thus
photon decay is a special case of the reaction $\gamma + \gamma_{\rm
CMB/EBL} \rightarrow e^++e^-$ with a vanishing $\gamma_{\rm
CMB/EBL}$.

Above the critical energy, with the energies of high energy photons
increasing, the threshold of low energy photons to interact also
increases. In the relevant energy range, the photon density of the
background lights decreases with energy. Hence, there are less low
energy photons to annihilate with high energy photons above the
critical energy, therefore, a recovery of the high energy photon
flux is expected in the LV theories above the critical
point~\cite{jp08}.

As for the photons with energies above EeV, Galaverni and Sigl
considered the ultra-high energy photons from GZK-cutoff-induced
pion decays~\cite{gs08,gs08b}. In their work, the LV parameters for
the electron and the positron are also fully taken under
considerations. For the non-LV case, these ultra-high energy photons
would interact with the CMB photons and lead to a cutoff in the
spectra (actually, a suppression here) above the pair-production
threshold. And if LV-corrected terms exist in the dispersion
relations, as shown above, the threshold will shift and there is a
reemergence above some critical energy if $\xi_n>0$. Hence, it is
supposed to observe the ``reemergent flux'' through Auger, AGASA,
and Yakutsk observations, according to numerical
calculations~\cite{gs08,gs08b}. However, these observations only
find upper limits for photon flux in the relevant energy range. The
non-observation situation allows very tight constraints on LV
parameters. Refs.~\cite{gs08,gs08b} reported that, because of the
non-observation of photons above $10^{19}~{\rm eV}$, $|\xi_1|$ is
limited to be below $10^{-14}$, and $\xi_2
> -10^{-6}$.
A recent work utilized the same scenario, intending to rule out the
space-time foam model mentioned above, after taking into the energy
loss of photons in the stochastic interactions with the
D-branes~\cite{mls10}.

\section{Constraints from Recent UHECRs Spectra and Discussions}

\label{sec4}

Due to the most violating processes and long baseline propagations
in astrophysics, there has been accumulating significant constraints
for LV parameters from the astrophysical society~\cite{kr08}. Aside
from the above mentioned scenarios, other processes are also taken
under considerations by many authors and have achieved good
progresses on the LV issues, {\it e.g.}, vacuum birefringence,
vacuum ${\rm\check{C}}$erenkov radiation, helicity decay, photon
splitting~\cite{jlm06,lm09,m05,kr08}.

The LV modifications are usually suppressed by the ratio of the
particle energy to the Planck scale, hence the high energy particles
have a merit to amplify these tiny effects. We here review some most
recent limits from UHECRs, because they possess the highest energies
we have ever observed, and can cast tight constraints on LV
theories.

UHECRs are among the most studied phenomena of LV for historical and
practical reasons. In a certain period of time, people doubted
whether a ``standard'' GZK cutoff exists, and many models are
proposed to account for the ``shift'' or ``disappearance'' of the
GZK cutoff. Among them, the LV explanation was a promising
candidate.

Nowadays, since the GZK cutoff is found around the predicted point,
it is used to constrain the LV
parameters~\cite{m09,b09,alfaro05,alfaro05b,xm09b,bi08,ss09}. Bi
{\it et al.} got the difference of the maximal attainable
velocity~\cite{cg99} of pions to photons, $\delta_{\pi} =
-0.8^{+3.2}_{-0.5} \times 10^{-23}$ and $\delta_{\pi} =
0.0^{+1.0}_{-0.4} \times 10^{-23}$, by utilizing the HiRes monocular
spectra and the Auger combined spectra, respectively, under the
assumptions that the LV modifications for nucleons are negligible
and that the dominant component of UHECRs is the proton~\cite{b09}.
Stecker and Scully derived a best fit to the LV parameter of
$3.0^{+1.5}_{-3.0} \times 10^{-23}$ from studies on the Auger
spectra~\cite{ss09}. The above two studies both refer to the
conception of the center of mass system, which might not be a
well-defined conception in presence of LV. Maccione {\it et al.}
worked in the framework of EFT with dimension-5 and dimension-6
operators, and utilized the Auger data to establish two-sided bounds
on the coefficients for the proton and the pion, whose LV scales are
found to be well above the Planck energy~\cite{m09}.

Though various UHECRs spectra studies favor the standard GZK
explanation, the LV scenarios are not ruled out thoroughly. Small
derivations from Lorentz symmetry are still under serious
considerations. However, the fact that the cutoff point together
with the spectrum of UHECRs are consistent with the standard
theoretical expectation favors the GZK explanation without LV. Some
authors argued that, the compound of UHECRs is not well understood
yet. Even for general knowledge, we believe that they are protons or
heavy ions. Then UHECRs are composite entities instead of
fundamental particles, and their underlying degrees of freedom are
actually quarks and gluons. This situation may limit the constraints
derived from UHECRs~\cite{gm04}. Moreover, the GZK cutoff may also
be due to the accelerating capability of the UHECRs host galaxies.
Therefore, more works on UHECRs from observational and theoretical
aspects are needed to address these related issues.

On the other hand, as mentioned, Refs.~\cite{gs08,gs08b} realized
that UHECRs lead to photo-pion production when undergoing inelastic
collisions with CMB photons, and subsequently pions decay into the
ultra-high energy photons, {\it i.e.}, the process $\pi \rightarrow
\gamma$ always accompanies with the GZK cutoff process $p +
\gamma_{\rm CMB} \rightarrow N + \pi$. After a research into the
influence of LV terms in this scenario, they reached the conclusion
that $|\xi_1|$ is limited to be below $10^{-14}$, and $\xi_2
> -10^{-6}$, in order to explain the non-observation of photons
in the UHECRs flux above $10^{19}~{\rm eV}$~\cite{gs08,gs08b}.

Ending this section, let us compare and discuss some LV scenarios.
The lag caused by the time-of-flight~\cite{a98,xm09} is the cleanest
test ever, and it introduces no interaction into the
phenomenological analysis, and some primary studies are emerging
with attempt to disentangle the unknown (thus controversial) source
effects~\cite{e99,e06,sxm10}. Hence, it provides the most convincing
evidence for LV effects on the photon sector. Time-of-flight
analysis favors linear energy dependence within current
observations. However, the results are contradictory with other
constraints from kinematics of reactions, where the linearly
corrected LV terms are constrained to be very
unlikely~\cite{jlm06,lm09,m05,kr08}. Even the stringy space-time
foam model, where no corrections for charged particles are predicted
thus avoiding many constraints from charged species, is debatably
unfavored by the non-observation of the ultra-high energy photon
flux~\cite{gs08,gs08b,mls10,emn10}. As for species other than
photons, {\it e.g.}, electrons, positrons, quarks, and gluons, they
can enjoy their own individual LV parameters differing from those of
photons in principle. In some theories, particles with different
helicities can also possess unrelated LV parameters. Thus lots of LV
parameters are involved totally, and detailed analysis is needed and
fortunately is performing to constrain all these tiny LV
terms~\cite{kr08}. Since most reactions contain more than one
species, combined analysis is severely in demand. As we can see from
the photon decays in the dimension-6 CPT-even EFT, the LV physics
generally depends highly, sometimes even nonlinearly, on the phase
space of LV parameters. Many efforts are still focusing to combine
constraints from different types of analysis to derive a robust
result~\cite{jlm06,lm09,m05,kr08}. As for composite particles, the
jump from fundamental degrees of freedom to the compounded entities,
{\it e.g.}, the protons and the heavy ions in the UHECRs, requires
more careful arguments and treatments.

\section{Summary}

\label{sec5}

Lorentz violation (LV) may arise from many quantum gravity (QG)
models. And apart from its origin, it is apparently important for
its own sake. Theoretically, LV physics can originate from stringy
considerations, effective field theories (EFTs), doubly special
relativity, loop gravity theories, and so on. Correspondingly,
fruitful and intriguing phenomena emerge from LV corrections, and
fortunately they are testable on several operating platforms from
various aspects nowadays, {\it e.g.}, the Fermi satellite and the
Pierre Auger Collaboration. This issue becomes more and more
attractive to the scientific society nowadays, because evidence and
constraints from experiments and observations are appearing
continuously and steadily reachable and reliable. The most promising
standpoint of the testability comes from the modified
energy-momentum dispersion relation, where extra LV terms are
introduced through different theoretical scenarios, compared to the
canonical energy-momentum relation, $E^2=p^2+m^2$.

In this paper, we extensively review the studies of the LV influence
on the astrophysical photons with very high energies. They are
classified into three categories: (i) the velocity of photons $v =
\partial \omega /
\partial k$ now depends on the wavelength of the light, hence introduces
a time lag between photons with different energies, which can be
observed from $\gamma$-ray bursts (GRBs) and the flares of active
galactic nuclei (AGNs); (ii) newfangled reactions that are strictly
forbidden in the standard model due to the non-conservation of
energy-momentum might happen within LV theories, and a cutoff of
photon flux above the threshold energy of photon decay, $\gamma
\rightarrow e^++e^-$, is expected with proper LV parameters; (iii)
LV also alters the threshold energy of normal reactions, hence the
spectra of the very high energy photons can be renewed to provide
physically intriguing phenomena, due to the modified patterns of the
reaction $\gamma + \gamma_{\rm CMB/EBL} \rightarrow e^++e^-$, for
multi-TeV photons and EeV photons through interacting with low
energy background lights.

To search for the LV physics in above scenarios, there are several
indispensable steps. Most importantly, more sensitive detectors
should play a decisive role, since generally, the higher the energy,
the more diluted the flux. Currently, the Auger Collaboration and
the HESS Collaboration are playing significant importance in
relevant fields. An ``Auger North'' array intending to behave better
than the present southern hemisphere Auger array is also under
consideration. An improved knowledge of the astrophysically
intrinsic mechanisms is also essential in searching LV physics from
astrophysical side, {\it e.g.}, the details of the background
lights, the un-attenuated spectra of sources, the acceleration and
radiation mechanisms of UHECRs. Space-based satellites, Fermi and
the incoming ``Extreme Universe Space Observatory onboard Japanese
Experiment Module'' (JEM-EUSO) are promising missions to pindown
astrophysical issues. As we can see, LV physics reveals itself at
high energy bands more obviously, thus the techniques to detect
higher energy cosmic rays, {\it e.g.}, $>10^{20}$~eV, are also of
great significance.

In summary, theoretically, we cannot predict the values of LV
parameters in LV theories from the first principle most of the time
though, fortunately, experimental and observational data are serving
tight constraints from various aspects to falsify theories and/or to
constrain the LV parameters. These observations have developed and
are developing rapidly within new platforms nowadays. Until now, no
explicit LV effects deviating from the special relativity are
observed determinately, and many theoretical scenarios are ruling
out. Meanwhile, new clues of LV physics are emerging rapidly from
astrophysical observations and ground-based laboratory experiments.
In the following dozens of years, the LV issue is expected to meet a
more clear situation then.

\begin{acknowledgments}
We thank helpful conversations with Zhi Xiao. This work is partially
supported by National Natural Science Foundation of China
(Nos.~11005018, 10721063, 10975003, 11035003). It is also supported
by Hui-Chun Chin and Tsung-Dao Lee Chinese Undergraduate Research
Endowment (Chun-Tsung Endowment) at Peking University, and by
National Fund for Fostering Talents of Basic Science (Nos.~J0630311,
J0730316).
\end{acknowledgments}

\end{document}